\documentstyle[preprint,eqsecnum,aps]{revtex}

\newcommand{\be}{\begin{equation}}
\newcommand{\eq}{\end{equation}}
   
\def\ov{\over}
\def\non{\nonumber }
\def\beq{\begin{equation} }
\def\eeq{\end{equation} }
\def\beqa{\begin{eqnarray}}
\def\eeqa{\end{eqnarray}}
\def\del{\partial }
\def\a{\alpha }

\begin{document}
\draft
\preprint{\parbox{1.5in}{SISSA 8/99/EP\\ 
           {\tt hep-th/9901092}}}
\title{Fundamental Strings as  Black Bodies}
\author{D. Amati{}$^a$ and J.G. Russo{}$^b$}

\address{${}^a$~International School for Advanced Studies (SISSA),
I-34013 Trieste, Italy\\
INFN, Sezione di Trieste\\
{\tt amati@sissa.it} }

\address{${}^b$~Departamento de F\'\i sica,  Universidad de Buenos Aires,\\
Ciudad Universitaria, Pabellon I, 1428 Buenos Aires.\\
{\tt russo@df.uba.ar} }

\maketitle
\begin{abstract}
We show that the decay spectrum of massive excitations in perturbative
string theories is thermal when averaged over the (many) initial
degenerate states. We first compute the inclusive photon spectrum
for open strings at the tree level showing that
a black body spectrum with the Hagedorn temperature emerges in the averaging.
A similar calculation for a massive closed string state with
winding and 
Kaluza-Klein charges
shows that the emitted graviton spectrum is thermal
with a ``grey-body" factor, which approaches one near extremality.
These results   uncover  a simple physical meaning
of the Hagedorn temperature and  provide an explicit  microscopic derivation
 of the black body spectrum from a unitary $S$ matrix.

\end{abstract}

\section{Introduction}


High string excitations have a large degeneracy; the number 
of string states of a given mass
increases exponentially in the excitation mass measured in string units (string
tension or inverse string length).  This degeneracy is consistent with the
unitarity of the perturbative $S$ matrix. 
Unitarity guarantees that the analysis of final states that arise by
the decay of a massive state must allow
to retrieve the information on how the original massive state has been formed, i.e.,
on which is the coherent superposition of the many degenerate
microstates that have been formed and subsequently decayed into the final
states under analysis.
A consistent and complete way of analysing final states is through
semi-inclusive quantities, such as spectra, two particle correlations, three
particle correlations, etc.  The total set of these  quantities
is complete in the sense that it characterizes the initial state or,
equivalently, the coherent superposition that created the final quantities
in question.  Semi-inclusive quantities are calculable in perturbative
string theory (order by order in the string coupling) with a well-defined 
algorithm \cite{uno},
which was extensively used in the past when string theory was being
applied to strong interactions.

In this paper a remarkable property of the decay spectrum of
a massive string state will be found: when
averaging over all string excitations of mass $M$ ($M^2\gg $~string~tension), the decay
spectrum exactly averages to a black body spectrum with Hagedorn temperature $T_H$.  
That the average procedure should wash out all the information stored in spectra and correlations leading to a total informationless thermal distribution was conjectured \cite{amati} in order to understand recently uncovered connections between string theory and black holes. 
For completeness we will briefly mention in section IV the possible relevance for black hole physics of this non-trivial property of a unitary microscopic quantum theory generating thermal properties in a decoherence procedure. 
The present results also unveil an important physical meaning of the Hagedorn temperature, as  the radiation temperature of a macroscopic (averaged) string.


\section{Decay rates in open string theory}

The inclusive spectrum for the decay of a particular 
massive state $|\Phi \rangle $ of momentum $p_\mu = (M,\vec 0)$ into a 
particle with momentum $k_\mu $ ($k^2=0$) is represented by the modulus squared of the 
amplitude for the decay of $|\Phi \rangle $ into a state $|\Phi_X\rangle\otimes 
| k_\mu \rangle $, summed over all  $|\Phi_X \rangle $. In string theory, the 
initial state $|\Phi \rangle $ will be described by a specific excitation
of the string oscillators (for simplicity, we restrict the discussion to 
bosonic string
theory).
If  $N_n$ is the occupation number of the $n$-th mode 
(with frequency $n$), a state $| \Phi \rangle $ of mass $M$ will be characterized 
by a partition $\{N_n\} $ of $N=\sum_{n=1}^\infty n N_n $, with 
\beqa
&&\hat N \ | \Phi _ {\{N_n\} } \rangle =N\ | \Phi _ {\{N_n\} } \rangle   
\ ,\ \ \  \hat N=\sum_{n=1}^\infty na_n^\dagger \cdot a_n \ ,
\non\\
&& \a'M^2= N-1 , \ \ \ [a_n^\mu , a_m^{\nu \dagger}]=\delta_{nm}\eta^{\mu\nu} , 
\label{uno}
\eeqa
where $\mu,\nu =1,...,D.$ For large $N$, there are \cite{ami,gsw} 
\beq
{\cal N}_N={\rm const.} \ N^{- {D+1\over 4}   } e^{a\sqrt{N} }\ ,\ \ 
\ \ \ a=2\pi\sqrt{D-2\over 6}\ ,
\label{numer}
\eeq
possible partitions, thus  representing the degeneracy of the 
level $N$ associated with excitations of mass $M$.
%
The total momentum $p_\mu '$ of $\Phi_ X$ will be
$p_\mu '=p_\mu - k_\mu $. 
At the tree level, $|\Phi_X\rangle $ will also be a superposition 
of string excitations, satisfying
$\hat N \ |\Phi_X\rangle  =N'\ |\Phi_X\rangle $, 
\beq
N'= \alpha '{p'}^2+1=N-2k_0\sqrt{\alpha' (N-1)}\ ,
\label{doss}
\eeq
Let us momentarily ignore twists or 
non-planar effects of open string theories
\cite{ami,gsw}. They will be incorporated later.
The inclusive photon spectrum for the decay of 
a state $\Phi_{ \{ N_n\} } $  is then given by 
\beqa
&&d\Gamma_{\Phi _ {\{N_n\} } }(k_0)
= \sum_{ \Phi_X|_{N'}} \big| \langle \Phi_X |
 V(k)|\Phi_{ \{ N_n \} } \rangle \big|^2
\non\\
&\times & V(S^{D-2} ) k_0^{D-3} dk_0
\ ,\ \ V(S^{D-2} )={2\pi^ {{D-1\over 2} }\over 
\Gamma ( {D-1\over 2}  ) } ,
\label{cuatro}
\eeqa
where $ \sum_{ \Phi_X |_{N'}}$ means the sum over all states $\Phi_X $
satisfying the mass condition (\ref{doss}), and $V(k)$ is the photon 
vertex operator
\beq
 V(k)=\xi_\mu \del_\tau X^\mu (0) e^{ik.X(0)} ,\  \  \xi_\mu k^\mu=0 ,\   
k^2=0 ,
\non
\eeq
\beqa
X^\mu (0)&=& \hat x^\mu + i\sum_{n=1}^\infty {1\over\sqrt{n}} 
\big( {a_n}^\mu -    
{a^\dagger_n}^\mu \big)\ ,
\non\\ 
\del_\tau X ^\mu (0)&=& \hat p^\mu + 
\sum_{n=1}^\infty \sqrt{n} \big( {a^\dagger_n}^\mu +
{a_n}^\mu \big) ,  \ [\hat x^\mu ,\hat p_\nu]=i\delta_\nu^\mu ,
\nonumber
\eeqa
where $\xi_\mu $ is the photon polarization vector, and we have set $\a ' =1/2$.
The $\delta $-function of energy-momentum conservation, coming from the 
zero-mode part of the amplitude, has been factored out from
(\ref{cuatro}). 
Let us define a projector $\hat P_N$ over states 
of level $N$ as 
\beq
\hat P_N= \oint _C {dz\over z} z^ {\hat N-N}\ ,\ \ \ \ 
\hat P_N|\Phi_{N'}\rangle =\delta_{NN'} |\Phi_{N'}\rangle \ ,
\label{seis}
\eeq
where $C$ is a small contour around the origin.
Introducing $\hat P_{N'}$ in (\ref{cuatro}) we can convert the sum into a sum 
over all physical states $|\Phi_X \rangle $, and use completeness to write
\beq
F_{\Phi _ {\{ N_n\} } }=\oint_{C} {dz'\over z'} {z'}^{-N'}
 \langle \Phi_{ \{ N_n\} }  |V(-k){z'}^{ \hat N } V(k)| 
\Phi_{\{ N_n\}  } \rangle  ,
\label{dosiet}
\eeq
\beq
F_{ \Phi _ {\{ N_n\} } }(k)
\equiv {   k_0^{3-D}   \over V(S^{D-2}) }{d\Gamma_{\Phi _ {\{N_n\} } }\over 
dk_0}\ .
\label{zino}
\eeq
Let us now introduce a coherent set basis,
\beq  |\{ \lambda_n\} \rangle=\prod_{n=1}^\infty \exp   
\big[\lambda_n\cdot a_n^\dagger \big] |0;p_\mu\rangle \ ,
\label{zinn}
\eeq
where $\lambda_n,\ n=1,..\infty $, is a set of arbitrary (complex) parameters,
and define
\beqa
F_{\lambda ^*\lambda }&=& \oint_C {dz\over z} z^{-N  }\oint _{C'}
 {dz'\over z'}{z'}^{-N ' } 
\non\\
&\times &\langle\{\lambda_n\} | W(-k){z'}^{\hat N  } 
W(k) z^{\hat N  }| \{\lambda _n\} \rangle ,
\label{zint}
\eeqa
where $W(k)\equiv\exp [\xi  \cdot   \del_\tau X^\mu  ]\exp [{ik.X}] $.
From standard properties of coherent states \cite{ami,gsw}, 
we find
\beqa
F_{\lambda ^*\lambda } &=& \oint_C {dz\over z} z^{-N  }\oint _{C'} {dz'\over z'}
{z'}^{-N ' }e^{p\cdot(\xi_1+\xi_2)} \prod_{n=1}^\infty \exp\big[ K_n
\non\\
&+& \lambda_n\cdot \lambda^*_n z^n {z'}^n
+\lambda^*_n\cdot J_n+ \lambda_n\cdot J_n^*\big]\ ,
\label{cohe}
\eeqa
\beqa
J_n^\mu &=& \sqrt{n} \big( \xi_2^\mu +\xi_1^\mu {z'}^n \big)-{1\over\sqrt{n} } 
k^\mu \big(1- {z'}^n \big)\ , 
\non\\
J_n^{*\mu } &=& \sqrt{n} \big( \xi_2^\mu  z^n {z'}^n  +
\xi_1^\mu {z}^n \big)-{1\over\sqrt{n}} k^\mu z^n \big(1-   {z'}^n  \big)\ ,  
\non\\
K_n &=&  \xi_1\cdot\xi_2 \  n {z'}^n \ .
\label{jotas}
\eeqa
The spectrum $F_{\Phi_{\{ N_n\} } }(k)$ may be directly computed from 
(\ref{dosiet}) using standard operator techniques or from the explicit expression (\ref{cohe}), (\ref{jotas}), and (we omit Lorentz indices)
%
%
%
%
%
%
\beq
F_{\Phi_{\{ N_n\} } }(k)=\prod _{n=1}^\infty \left( {\partial\over\partial \lambda_n }\right) ^{N_n} 
 \left( {\partial\over\partial \lambda_n^*} \right) ^{N_n} 
F_{\lambda ^*\lambda } \bigg|_{ {
{\rm linear\ in\ }\xi_1\xi_2\ \atop \lambda_n=\lambda^*_n=0} }\ ,
\eeq
which follows after inverting  (\ref{zinn}) and noticing that $V(k)$ is represented by the linear term in $\xi $ of $W(k)$.
For every $|\Phi_{\{ N_n\} }  \rangle $, the $F_{\Phi_{\{ N_n\} } }(k)$
are easily obtained and are polynomials in~$k_\mu $.

The spectrum obtained 
by averaging over string states with mass $M$ is 
\beq
F_{N}(k_0)={1\over 
{\cal N}_N} \sum_{\Phi|_N} F_{\Phi_{\{ N_n\} } } (k_0)\ .
\label{qql}
\eeq
One can introduce again the projector (\ref{seis}) 
to convert the above sum into a sum over
 all states, so that 
\beqa
F_{N}(k_0)&=&{1\over {\cal N}_N}\oint_C {dz\over z} z^{-N  }  \oint _{C'} {dz'\over z'} 
{z'}^{-N ' }  
\non\\
&\times & {\rm Tr}\ \big[z^{ \hat N } V(-k){z'}^{ \hat N } V(k)\big] \ .
\label{zin}
\eeqa
The trace in (\ref{zin}) can then be computed from (\ref{cohe}) by integrating over  
$\lambda _n,\lambda_n^* $,
\beq
F_N(k_0)={1\over {\cal N}_N}\int \prod_{n,\mu} d\lambda_n^\mu 
d{\lambda_n^*}^\mu e^{-\lambda_n^*\cdot \lambda_n } 
F_{\lambda ^*\lambda }    \bigg| _{\xi_1\xi_2 }  ,
\label{ocho}  
\eeq
where 
$\big| _{\xi_1\xi_2 } $ means
the term  linear  in $\xi_{1\mu} ,\xi_{2\nu } $, and then 
setting $\xi_{1\mu} =\xi_{2\mu} $.
Using (\ref{cohe}) we get
\beqa
F_{N}(k_0)&=& {e^{p\cdot(\xi_1+\xi_2)}\over {\cal N}_N}
 \oint_{C_1} {dw\over w} w^{-N} f(w)^{2-D}\oint _{C_2}
 {dv\over v}  v^ {N-N'} 
\non\\
&\times &\ \exp\big[ J_n 
\cdot J^*_n(1-w^n)^{-1} +K_n\big] \ \big| _{\xi_1\xi_2} ,
\non\\
f(w)&=&\prod_{n=1}^\infty (1-w^n)\ ,\ \ \ \ w=zz'\ ,\  \ v=z'. 
\label{deed}
\eeqa
$C_1$ and $C_2$ are two small contours around zero.
The extra factor $f^2(w)$ in the integrand arises as usual by incorporating 
the ghost contribution (the  $D-2$ power corresponds to the $D-2$ transverse 
coordinates). Thus we obtain
\beqa
F_{N}(k_0)&=&{1\over {\cal N}_N}\xi_{\mu}\xi_{\nu}\oint_{C_1} 
{dw\over w} w^{-N} f(w)^{2-D} \oint _{C_2} {dv\over v}  v^ {N-N'}
\non\\
&\times & 
\left( p_\mu p_\nu+\eta_{\mu\nu} \Omega (v, w)\right) \ ,
\label{zinzz}\\
\Omega(v,w)&=& 
\sum_{n=1}^\infty n \bigg( v^n +   {w^n(v^n+v^{-n})\over 1-w^n} \bigg) \
 .
\label{yome}
\eeqa
The integral over $v$ gives ($N'<N$)
\beq
F_{N}(k_0)={\xi ^2 \ (N-N') \over {\cal N}_N} \oint 
 {dw\over w} {w^{-N'} f(w)^{2-D}  \over 1- w^{N-N'} }  .
\label{wwww}
\eeq
For large $N'$, the integral in (\ref{wwww}) can be computed by a saddle point 
approximation, with the main contribution coming from $w\sim 1$.
This is similar to the familiar calculation of the number of states 
${\cal N}_N$ (\ref{numer}) of  level $N$. 
The behavior of $f(w)$ at $w\cong 1$  is well known and given by
\beq
f(w)\cong {\rm const. }\ 
(1-w)^{-1/2} e^{-{\pi ^2 \over 6 (1-w)} }\ .\ \
\label{deede}
\eeq
Therefore the saddle point is at $\ln w \cong -a/ 2\sqrt{N'}$. 
Using $N'=N-2k_0\sqrt{\a' N}$ we obtain for large $N'$
\beq 
F_{N}(k_0)\cong \xi ^2 \ 2k_0\sqrt{\a' N} \ {e^{a\sqrt{N-2k_0\sqrt{\a 'N} } -a\sqrt{N} }
 \over 1-e^{-a k_0\sqrt{\a'}} }\ .
\label{zzz}
\eeq
{}We have set 
$(N/N')^{{D+1\over 4}}\cong 1$,
since
 in the regime of interest,  $k_0\ll M$,
$N/N'\cong 1$ (the emission of a photon with energy 
$k_0$ of order $M$
and higher is suppressed by a factor of order $e^{-\sqrt{N}}$). 
Thus $\sqrt{N'}-\sqrt{N}\cong -k_0\sqrt{\a'} $ 
and
\beq 
F_{N}(k_0)\cong \xi ^2 2\a '  k_0 M
 \ {e^{-ak_0\sqrt{\a '}} \over 1-e^{-a k_0\sqrt{\a'}} }\ ,
\label{zzzh}
\eeq
or
\beq
d\Gamma_N(k_0)\cong {\rm const.} \ 
{e^{-{k_0\over T_H}} \over 1-e^{- {k_0\over T_H}}  }\ k_0^{D-2} dk_0\ ,
\label{hzzh}
\eeq
where 
$T_H$ is the Hagedorn temperature $T_H={1\over a\sqrt{\a' }}$. 
Thus the radiation 
spectrum from a macroscopic (i.e. averaged over all degenerate microscopic
states) string 
of mass $M$
is exactly thermal (despite the spectrum for each microstate being absolutely
non-thermal).
The way the exact Planck formula arises in the final result is striking, since
the exponents in the numerator and denominator have different origins.


Let us now show that the non-planar contribution does not change the above result.
It arises in open string theory due to the fact that the photon may be emitted by any of the two ends of the string.
So, strictly speaking, we should have written
\beqa
&&d\Gamma_{\Phi _ {\{N_n\} } }(k_0)
= \sum_{ \Phi_X|_{N'}} \big| \langle \Phi_X |
 {1\over\sqrt{2}}(V(k)+\Theta V(k)\Theta ) |\Phi_{ \{ N_n \} } 
\rangle \big|^2
\non\\
&\times & V(S^{D-2} ) k_0^{D-3} dk_0
\ ,\ \
\label{zuatro}
\eeqa
at the place of (\ref{cuatro}), where $\Theta $ is the twist operator  
\cite{ami}.
The modulus squared involving $V(k)$ --and that with $\Theta V(k)\Theta $--
leads each to half of the planar result (\ref{hzzh}) obtained before.
The cross product gives rise to the non-planar contribution
that may be computed analogously
to the planar loop diagram, of which our spectrum represents the absorptive part.
Instead of (\ref{zin}) for the non-planar contribution one has
\beqa
F_{N}(k_0)_{\rm n.p.}&=&
{1\over {\cal N}_N}\oint_C {dz\over z} z^{-N  }  \oint _{C'} {dz'\over z'} 
{z'}^{-N ' }  
\non\\
&\times & {\rm Tr}\ \big[z^{ \hat N }\Theta V(-k)\Theta {z'}^{ \hat N } V(k)\big] \ .
\label{yin}
\eeqa
Using the explicit expression for $\Theta $ \cite{ami} 
and computing the trace  \cite{gross}
 one finally obtains
\beqa
F_{N}(k_0)_{\rm n.p.} &=&{1\over {\cal N}_N}\xi_{\mu}\xi_{\nu}\oint_{C_1} 
{dw\over w} w^{-N}(1-w) f(w)^{2-D} \oint _{C'} {dz'\over z'}  {z'}^ {N-N'}
\non\\
&\times & 
\left( p_\mu p_\nu+\eta_{\mu\nu} \Omega (v, w)\right) \ ,
\label{yiny}
\eeqa
where $\Omega $ is given by eq.~(\ref{yome}) with
$v={w-z'\over 1-z'}$.
The contour integral in $z'$ now selects the pole at $z'=w$. 
The remaining integral over $w$ is dominated by the same saddle point
 that allowed the evaluation
of (\ref{wwww}). The final result is
\beq
F_{N}(k_0)_{\rm n.p.} \cong {\xi^2 a\over 2\sqrt{N} } e^{-ak_0\sqrt{\a '} }
\eeq
which is smaller by  a factor $< 1/N$ with respect to the planar result 
(\ref{zzzh}) or (\ref{hzzh})
that dominates the sum (\ref{zuatro}).

\section{Spectrum of a charged closed string state}

Let us now consider closed bosonic string theory with one dimension 
compactified on a circle of
radius $R$. Let $m_0$ and $w_0$ be the integers representing the 
Kaluza-Klein momentum and winding number of the
string state along the circle (we will assume $m_0,w_0>0$).
The mass spectrum is given by
\beq
\a'M^2=4(\hat N_R-1)+ \a' Q_+^2= 4(\hat N_L-1)+\a' Q_-^2\ ,
\label{aggg}
\eeq
where $Q_\pm ={m_0\over R}\pm {w_0R\over\a'}$, 
Consider the decay rate of a state of level $N_L,N_R$ of given
charges $m_0,w_0$, with 
$N_L-N_R=m_0w_0$, by
averaging over all physical states of charge $m_0,w_0$ and level $N_L,N_R$.
The calculation factorizes in a left and right part, and it is similar
as in the previous section. We assume that $N'_R$ and $N'_L$ are sufficiently 
large for the saddle-point
approximation to apply. 
The final result is
\beqa
d\Gamma_{N_RN_L}(k_0) &=& {\rm const.}\ \xi ^2 M^2\
 {e^{-{ k_0\over T_L} } \over
1-e^{- {k_0\over T_L }} } \ {e^{-{ k_0\over T_R} } \over
1-e^{- {k_0\over T_R}} }
\non\\
&\times & V(S^{D-2}) k_0^{D-1}  dk_0  \ ,
\label{qzzz}
\eeqa
\beq
T_R={2 \sqrt{M^2-Q_+^2} \ov a \sqrt{\a' } \ M}, 
\ \ \ T_L= {2\sqrt{M^2-Q_-^2} \ov a \sqrt{\a' }\ M}\ .
\label{tempe}
\eeq

Equation (\ref{qzzz}) is remarkably similar to the analogous result derived for D-branes in 
\cite{CM} (see also \cite{malstro} ), despite the two calculations
being rather different, and  describing  different physical systems.
Indeed, the D-brane calculation involves only
a single oscillator $a,a^\dagger$ that excites modes with one unit of
 Kaluza-Klein momentum \cite{CM};  
the present elementary string calculation
 involves all excitations
$a_n,a_n^\dagger,\ n=1,...,\infty$, in a 
very specific way dictated by the vertex insertion.

By analogy with black holes  \cite{malstro},
eq.~(\ref{qzzz})  may be written as
\beq
d\Gamma_{N_RN_L}(k_0) ={\rm const. }\ \sigma(k_0)\  
 {e^{-{ k_0\over T} } \over
1-e^{- {k_0\over T }} } \ k_0^{D-2} dk_0\ ,
\eeq
where $T^{-1}=T_R^{-1}+T_L^{-1}$ and 
$\sigma(k_0) $ is the ``grey body" factor
\beq
\sigma (k_0)= k_0 \ { 1-e^{-{k_0\over T} } \over  (1-e^{- {k_0\over T_L}}) 
(1-e^{- {k_0\over T_R}}) }   \ .
\label{dzzz}
\eeq

The radiation vanishes if the initial string state saturates
 the BPS condition $M\geq 
Q_+$, i.e.
when $ M=Q_+ $. For a  near BPS state, $M\cong 
Q_+$, $N_L\gg N_R $,
one gets
\beq
T\cong T_R= {2\over a \sqrt{\a' }}{\sqrt{M^2-Q_+^2} \ov M}\ .
\label{nebps}
\eeq
In the superstring theory,  
owing to the supersymmetry of the BPS state, corrections to the 
``free string" picture are expected to be
smaller for near-BPS configurations, which might allow to extrapolate
eq.~(\ref{nebps}) 
from the weak  to the strong coupling  regime \cite{cuerda}.


\section{Discussion}

The appearance of the Hagedorn temperature characterizing the radiation
 in the decay of a massive macrostate in string theory 
at zero temperature 
sheds light on an old problem concerning the physical interpretation of 
the Hagedorn temperature.
The Hagedorn temperature was traditionally  interpreted
as the temperature at which the canonical thermal ensemble
of a string gas breaks down. The technical reason is simple  
--at  higher temperatures the integral defining the thermal partition function 
diverges-- and it is related to the fast growing of the level density with mass.
The question is what is the physical picture
behind this instability. 
The present results lead to the following interpretation
(this seems consistent  with other suggestions \cite{turok,giddi}).
When an open string in a macrostate with $N\gg 1$ is 
placed in a thermal bath of 
temperature $T_{\rm bath}<T_H$,  
the string will decay by emitting massless (as well as massive) 
particles,
as a black body of temperature $T_H$, until the level  decreases to a point where 
the saddle-point prediction $T_{\rm string}=T_H$ ceases to apply,
and the real temperature of the string is of order $T_{\rm bath}$. 
A situation of equilibrium occurs immediately if 
one places the string in a thermal bath with $T_{\rm bath}=T_H$.
When $T_{\rm bath} >T_H$, the energy of the string  increase endlessly, 
since it can only emit at
a fixed rate determined by the temperature $T_H$. The system becomes highly unstable, 
and there cannot be any thermal equilibrium. The
 canonical thermal ensemble must therefore break down precisely at $T=T_H$.


It is interesting to witness how these statistical or classical concepts stem from microstate computations through average procedures (decoherence). 
The string scale is of course present
in the spectrum and in the correlations of single microstates,
but it is only after averaging that appears as the temperature 
characterizing the emission. 
The emergence of classical properties in a computable weak coupling
string regime was indeed one of the
motivation for this investigation. It substantiates the conjecture 
\cite{amati}
that a similar mechanism is at work at large couplings in order to 
generate from string theory 
classical concepts such as space-time, causal relations and black hole physics.

Finally, it would  be interesting to investigate possible cosmological implications related to the decay spectrum and lifetime of cosmic strings \cite{turok,kibble}.

\acknowledgements{
J.R. would like to thank SISSA for hospitality. This work was partially supported by EC Contract ERBFMRXCT960090 and by the research grant on Theoretical Physics of Fundamental Interactions of the Italian Ministry for Universities and Scientific 
Research.

}

\end{document}